# Effects of grain boundary on the sources of size effects

**George Z. Voyiadjis, Mohammadreza Yaghoobi**
Computational Solid Mechanics Laboratory, Department of Civil and Environmental Engineering,
Louisiana State University, Baton Rouge, LA 70803, United States

**Abstract:** The present work investigates the effects of grain boundary (GB) on the sources of size effects. Up to now, several studies have been conducted to address the role of GBs in size effects from the atomistic point of view. However, a study which addresses the effects of GB on different governing mechanisms of size effects as the sample length scale varies has not been presented yet. The results show that the main role of GB at larger length scale is to change the pattern of dislocation structure in a way that the dislocations are piled up near the GB which increases the hardness.

**Keywords:** Grain boundary; Size effects; Molecular dynamics; Dislocation; Microstructures

**1.    Introduction**

The mechanical properties of crystalline metals are usually governed by grain size and grain boundary (GB) properties [1-3]. The GB effects on the strength depend on several factors including the grain size, GBs geometry, mechanical and crystallographical structure of metal, strain rate, and temperature [1-3]. These factors define the deformation mechanisms in crystalline metals. Several numerical and experimental studies have been conducted to unravel the role of GBs in crystalline metals [1-3]. In the case of large grains, i.e. grain sizes of more than 1 μm, the strength increases as the grain size decreases which is commonly described by the Hall-Petch relation. The enhancement in strength is justified using the dislocation pile-up mechanism [1-3]. By decreasing the grain size, the pile-up model and consequently the Hall-Petch relationship break down at some grain size which is of the order of nanometers [4]. Other mechanisms such as GB sliding, GB rotation, and GB dislocation creation and annihilation have been proposed to describe the grain size dependency of strength in nanocrystalline metals [1-3]. Many methods have been developed to study the mechanical and tribological properties of materials [5-7]. However, nanoindentation is very common to investigate the size effects in metallic samples of confined volumes [8-11]. In the current study, the effects of GBs on the sources of size effects are investigated. The nanoindentation response of single and bi-crystal Ni thin films are investigated by incorporating large scale atomistic simulation.

**2. Simulation details and methodology**

To investigate the effects of grain size on the sources of size effects, two different sizes of Ni thin films with the dimensions of 24 nm × 24 nm × 12 nm (S1) and 120 nm × 120 nm × 60 nm (S2) are generated and simulated using the classical molecular dynamics. The total number of atoms in the samples are around 650,000 for S1 and 79,000,000 for S2. The periodic boundary conditions are applied to the surrounding surfaces. The free boundary conditions are selected at the bottom of the sample. To prevent the whole domain from translational movement, at each step, the total force is divided by the number of atoms and then applied in the opposite direction to all atoms [8-11]. The velocity Verlet algorithm with a time step of 2 fs is selected. The parallel MD code LAMMPS is used [12]. A symmetric tilt boundariy of $\sum 3\ (1\ 1\ 1)\ [1\ \bar{1}\ 0]\ (\theta = 109.5°)$, is selected to study the effects of GB properties on the size effects. The GBs are located at a third of the thickness from the top surface. For each bi-crystal sample, one single crystal sample is also generated which has the atomic structure similar to that of the upper grain. Two spherical indenters with the radius of $R_1 = 10$ nm and $R_2 = 15$ nm are selected to perform indentation on S1 and S2 samples, respectively. The nickel embedded-atom method potential developed by Mishin et al. [13] is incorporated. The mean contact pressure $p_m$ can be obtained by deviding the indentation load $P$ by contact area $A$.

**3. Results and Discussions**

The results for the small (S1) and large (S2) sample in the cases of $\sum 3\ (1\ 1\ 1)\ [1\ \bar{1}\ 0]$ grain boundary is presented in Fig. 1. In the case of small sample, the size effects can be studied in five different regions of indentation depths:
*Region I*: Here, the only defect existing in bi-crystal sample is the $\sum 3\ (1\ 1\ 1)\ [1\ \bar{1}\ 0]$ GB. In the case of single crystal, the sample is defect-free. The response in this region is elastic for both bi-crystal and single crystal samples. There is no difference between the single crystal and bi-crystal samples responses in this region.
*Region II*: Here, the resulting mobile dislocation density is increased by increasing the total dislocation length. Consequently, the required applied stress to sustain the plastic deformation decreases. In other words, the dislocations created in the bicrystal sample reduce the strength. In the case of single crystal, the sample is still defect-free and the response remains elastic.
*Region III*: In the case of single crystal sample, the first dislocation is nucleated beneath the indenter which induces a large drop in hardness. In the case of the bi-crystal sample, the dislocation length remains nearly constant. According to the source exhaustion mechanism, the applied stress should increase to sustain the plastic flow which increases the strength.
*Region IV*: In *Regions III and IV*, the dominancy of source exhaustion mechanism decreases by increasing the total dislocation length in the case of single crystal sample. Consequently, the slope of hardness reduction decreases as the total dislocation length increases which is obvious in *Region IV*.
*Region V*: Increasing the indentation depth, the dislocation interaction with each other becomes important which activates the forest hardening mechanism. Also, the available dislocation length is enough to sustain the applied plastic flow. Consequently, the source exhaustion mechanism is not active anymore.

In the case of large sample with the $\sum 3\ (1\ 1\ 1)\ [1\ \bar{1}\ 0]$ grain boundary, the size effects can be studied in three different regions of indentation depths:
*Region I*: The response in this region is elastic for both bi-crystal and single crystal samples. Besides the GB, there are no other defects in both bi-crystal and single crystal samples.
*Region II*: The initial dislocation nucleation occurred beneath the indenter. A large drop in hardness occurs immediately after the first dislocation nucleation for both bi-crystal and single crystal samples. The size effects in bi-crystal and single crystal samples are controlled by dislocation nucleation and source exhaustion. As the total dislocation density increases, the resulting mobile dislocation density also increases which leads to the lower hardness according to the source exhaustion mechanism. In Region II, the difference between the total dislocation length of the bi-crystal and single crystal samples is small which leads to the nearly similar hardness for the bi-crystal and single crystal samples according to the source exhaustion mechanism. Increasing the indentation depth, the GB starts blocking the movement of dislocations.
*Region III*: As the total dislocation length increases, the source exhaustion mechanism becomes inactive in this region due to the fact that enough dislocation length is available to sustain the applied plastic flow. On the other hand, the forest hardening mechanism gradually becomes activated as the total dislocation length increases. The total dislocation length of the bi-crystal sample starts deviating from that of the single crystal sample, and it has larger values. Furthermore, the total length of dislocation blocked by the GB becomes noticeable. In Region III, the hardness in the bi-crystal samples becomes larger than that of the single crystal sample.

Also, as an example, Fig. 2 presents the dislocation structure in different regions for both single crystal and bi-crystal samples in the case of $\sum 3\,(1\,1\,1)\,[1\,\bar{1}\,0]$ GB in the case of large sample.

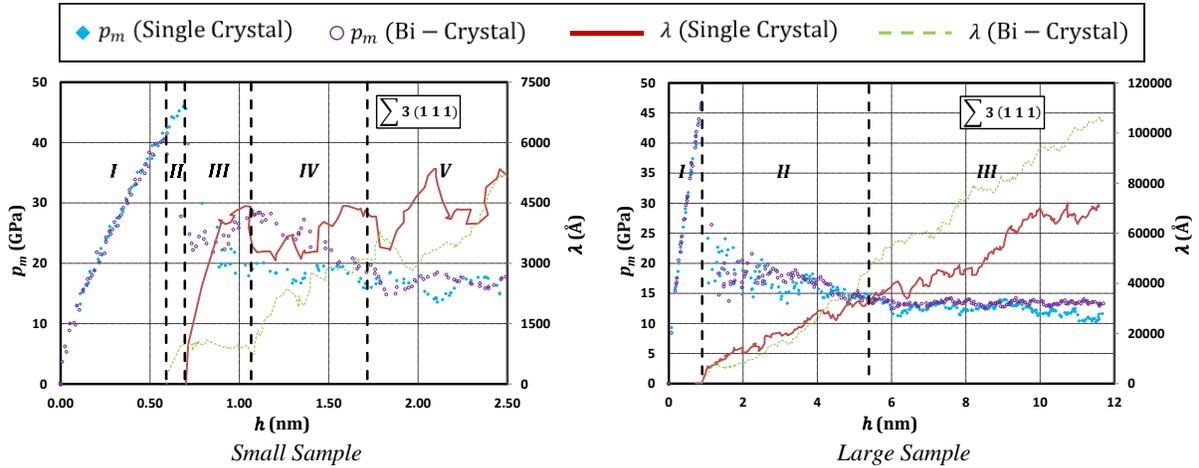

*Small Sample*          *Large Sample*

**Fig. 1.** Variation of mean contact pressure $p_m$ and dislocation length $\lambda$ as a function of indentation depth $h$ for small and large bicrystal samples with $\sum 3\,(1\,1\,1)$ GB and its related single crystal sample.

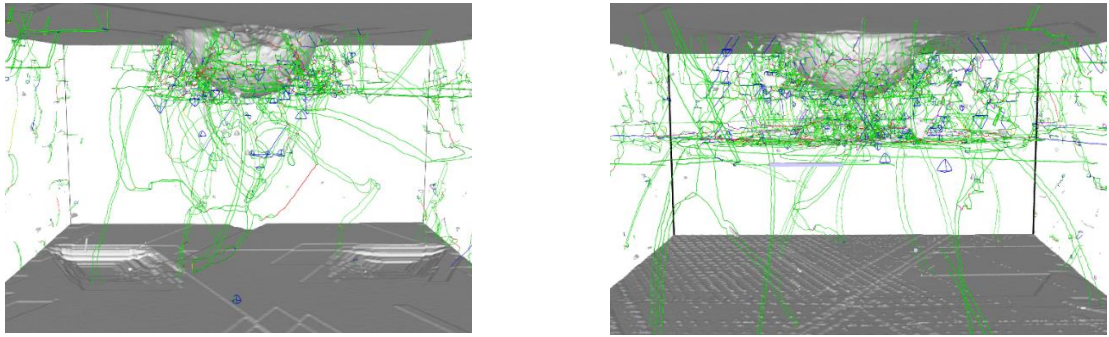

(a) single crystal sample, $h \approx 11.5$ nm        (b) bi-crystal sample, $h \approx 11.5$ nm

**Fig. 2**. Dislocation nucleation and evolution in large sample: (a) single crystal sample, $h \approx 11.5$ nm; (b) bi-crystal sample, $h \approx 11.5$ nm.

## 4. Conclusion

The effects of GB on the sources of size effects depend on the sample length scale. GB does not change the general pattern of size effects; however, it can contribute to some specific mechanisms depending on the sample size. The size effects is initially controlled by source exhaustion mechanism during the nanoindentation which is influenced by GB in the cases of small samples. The initial strength drop occurs earlier in small bi-crystal samples compared to their related single crystal samples. It may be originated from the fact the total dislocation lengths of bi-crystal samples are larger than those of the related single crystals at the moment of strength drop which ease the sudden hardness reduction according to the source exhaustion mechanism. The forest hardening mechanism becomes dominant at larger indentation depths. In the cases of small samples, however, the responses of bi-crystal and single-crystal samples are similar in this region. It shows that the GB does not change the nanoindentation response at larger indentation depths when the source exhaustion mechanism becomes inactive and forest hardening is dominant.

In the cases of large samples, the total dislocation content is much greater than that of the small samples. Hence, the interaction of dislocations and GBs plays a key role. The initial nanoindentation responses of bi-crystal and single crystal samples are similar in the cases of large samples. It is due to the fact that the total dislocation length of the bi-crystal sample is close to that of the single crystal sample. It means that the GB does not have any noticeable effect on the source exhaustion mechanism. The dislocations moving downward are blocked by the GB. At higher indentation depths, the strength of bi-crystal samples becomes larger than that of the single crystal samples. The total dislocation lengths are large at higher indentation depths, and consequently the density of dislocations piled-up behind the GB becomes noticeable. Hence, the increase in bicrystal samples can be justified according to the forest hardening mechanism.

## 5. Acknowledgments

The current work is funded by the NSF EPSCoR CIMM project under award #OIA-15410795. This research was conducted with high performance computing resources provided by Louisiana State University (http://www.hpc.lsu.edu) and Louisiana Optical Network Initiative (http://www.loni.org).